\documentclass[a4paper,fleqn,usenatbib]{mnras}

\usepackage{newtxtext,newtxmath}
\usepackage{morefloats}

\usepackage[T1]{fontenc}
\usepackage{ae,aecompl}

\usepackage{graphicx}	
\usepackage{amsmath}	
\usepackage{amssymb}	
\usepackage{float}
\usepackage{placeins} 
\usepackage{flafter}  

\title[The variability of the warm absorber in I~ZW~1]{The variability of the warm absorber in I Zwicky 1 as seen by XMM-{\em Newton}}

\author[C. V. Silva et al.]{
C. V. Silva,$^{1,2}$\thanks{E-mail: silva.cvj@gmail.com}
E. Costantini,$^{1,2}$\thanks{E-mail: e.costantini@sron.com}
M. Giustini,$^{2}$
G. A. Kriss,$^{3}$
W. N. Brandt,$^{4,5,6}$
\newauthor
L. C. Gallo,$^{7}$
and D. R. Wilkins$^{8}$
\\
$^{1}$Anton Pannekoek Institute for Astronomy, University of Amsterdam, Science Park 904, 1098 XH Amsterdam, The Netherlands\\
$^{2}$SRON, Netherlands Institute for Space Research, Sorbonnelaan 2, 3584 CA Utrecht, The Netherlands\\
$^{3}$Space Telescope Institute, 3700 San Martin Drive, Baltimore, MD 21218, USA\\
$^{4}$Department of Astronomy and Astrophysics, 525 Davey Lab, The Pennsylvania State University, University Park, PA 16802, USA\\
$^{5}$Institute for Gravitation and the Cosmos, The Pennsylvania State University, University Park, PA 16802, USA\\
$^{6}$Department of Physics, 104 Davey Lab, The Pennsylvania State University, University Park, PA 16802, USA\\
$^{7}$Department of Astronomy \& Physics, Saint Mary's University, Halifax, NS. B3H3C3, Canada\\
$^{8}$Kavli Institute for Particle Astrophysics and Cosmology, Stanford University, 452 Lomita Mall, Stanford, CA 94305, USA
}

\date{Accepted XXX. Received YYY; in original form ZZZ}

\pubyear{2018}

\begin{document}
\label{firstpage}
\pagerange{\pageref{firstpage}--\pageref{lastpage}}
\maketitle

\begin{abstract}
We present new XMM-{\em Newton} observations of the intriguing warm absorber in I~Zwicky~1. This luminous and nearby narrow-line Seyfert 1 galaxy shows ionized absorption by two components of outflowing gas; a low and a high-ionization phase with log $\xi\sim0$ and log $\xi \sim2$ respectively. Detailed modelling of these data reveal a complex and variable multi-phase warm absorber. However, we find the changes in the ionization state of the gas not to be straightforwardly correlated with the variability of the intrinsic continuum source, in apparent contrast with photoionization equilibrium. The observed variability hints instead at a close connection between the two gas components, possibly both directly connected to the accretion disc activity. We thus suggest a phenomenological model capable of explaining these observations, consisting of a clumpy outflow where the high and the low-ionization components are closely linked. Changes in ionization over the years are mainly driven by the different densities of the clumps crossing the observer's line-of-sight, in which the `skin' layer facing the source accounts for the more ionized component.
\end{abstract}




\section{Introduction}
Ionized outflows in Active Galactic Nuclei (AGN) have long been a subject of study in the X-ray and ultraviolet (UV) domains. Such gas is frequently referred to as a warm absorber (WA) and its presence is inferred from the detection of blueshifted absorption lines in the high-resolution spectra of these sources \citep[for a review see][]{crenshaw2003}. It has been estimated that roughly 60$\%$ of Seyfert 1 galaxies show the presence of a warm absorber in their spectra \citep[e.g.][]{crenshaw1999,laha2014}.\\
\indent Despite extensive studies, the origin and physical structure of warm absorbers is not yet fully understood. In the UV, the observed transitions belong mainly to a few important ions (e.g. \ion{C}{iv}, \ion{N}{v}, \ion {O}{vi}, and Lyman-$\alpha$), but thanks to the high spectral resolution available in this band several velocity components can be distinguished, while in the X-rays dozens of blurred transitions make it possible to estimate the ionization state and column density of the different gas components. Other important characteristics such as the spatial extent of the absorbing gas and its distance relative to the central source are harder to determine. The spatial location of the outflows in particular yields valuable information for AGN feedback studies \citep[e.g.][]{dimatteo2005,hardcastle2007,crenshaw2012,fabian2012}. An estimate of the density $n$ of the gas grants an estimate of its distance $r$ to the central source since the ionization parameter $\xi$ is a function of both these properties, as well as of the ionizing luminosity $L_\text{ion}$, $\xi=L_\text{ion}/nr^{2}$. In the UV, sensitive absorption lines from meta-stable transitions can be used to measure the density of the gas \citep[e.g.][]{kraemer2006,arav2008,arav2015}. In the X-rays, the density can be estimated through variability studies. The intrinsic X-ray source is variable and changes in the ionizing flux induce a response in the ionization state of the gas, characterized by an equilibrium timescale. The time it takes for the gas to reach equilibrium with the ionizing continuum is dependent on the properties of the gas, specifically on its density \citep{nicastro1999,krolik&kriss2001,silva2016}. Obtaining the density, and subsequently the distance, of the warm absorber via this method requires detailed monitoring of the source and has been applied to several objects through time-resolved spectroscopy \citep[e.g.][]{krongold2007,steenbrugge2009} and time-dependent photoionization studies \citep{nicastro1999,kaastra2012}. However, some sources appear to show an even more complex behaviour. For example, in MR 2251-178 there seems to be no connection between the ionization parameter and the X-ray luminosity \citep{kaspi2004}. More recently, \cite{longinotti2013} reported the discovery of intrinsic ionized absorption in Mrk 335, for which no correlation was found between the warm absorber variability and the X-ray flux.\\
\indent Multiwavelength UV-X-ray campaigns are key to characterize the outflow \citep[for a review see][]{costantini2010}. Previous studies have suggested ionized absorption in the X-rays to be a manifestation of the same gas that absorbs in the UV \cite[see e.g.][]{kaspi2002,arav2007,ebrero2011}. Unveiling this connection between the UV and X-ray absorbing gas is fundamental to understanding the nature of warm absorbers and, consequently, assess the impact of the gas outflows on the surrounding environment.\\
\indent I Zwicky 1 (I~ZW~1) is a narrow-line Seyfert 1 galaxy located at a redshift of z=0.061169 \citep{springob2005}. Previous XMM-{\em Newton} observations of this source revealed absorption by two components of ionized gas and an apparent anti-correlation between X-ray ionization and ionizing luminosity on timescales of years \citep{costantini2007}. In this work, we further analyse the absorption-line rich soft X-ray spectrum of I~ZW~1 as recently observed by XMM-{\em Newton}.\\
\indent Details on these observations and data reduction are reported in section \ref{section:2}. Section \ref{section:3} refers to the detailed analysis of the data and modelling of the spectral features through time-averaged spectral fitting and time-resolved spectroscopy. Section \ref{section:4} is devoted to an extensive discussion of our results and possible physical scenarios for the nature of the warm absorber. The conclusions and a summary of our work can be found in section \ref{section:5}.\\
\indent We use a flat cosmological model with $\Omega_{m}=0.3$, $\Omega_{\Lambda}=0.7$, and $\Omega_{r}=0.0$, together with a Hubble constant $\text{H}_{0}=70\ \text{km}\ \text{s}^{-1}\ \text{Mpc}^{-1}$. For the spectral modelling in this paper we have assumed solar abundances \citep{lodders2009} and a Galactic column density of $N_{\text{H}_\text{tot}}=6.01\times10^{20}\ \text{cm}^{-2}$, which includes both the atomic and molecular hydrogen components \citep{elvis1989,willingale2013}. The errors quoted in this paper are $1\sigma$ errors unless otherwise stated.
\section{Observations and data reduction}
\label{section:2}
I~ZW~1 was observed by XMM-{\em Newton} during two consecutive orbits on January 21 and January 22, 2015 (hereafter observations 301 and 801, respectively) as part of a multiwavelength campaign (PI: Costantini, E.), also including simultaneous observations with HST. The analysis of the X-ray time variability are presented in \cite{wilkins17}. The broadband EPIC-pn spectra will be treated in a follow-up paper (Gallo et al. in prep.). Finally, the analysis of the HST data will be published in a subsequent paper by Giustini et al. (in prep.). Here we focus on the absorption-line rich soft X-ray spectrum of I ZW 1 observed with the RGS instrument on board XMM-{\em Newton}.\\
For these observations, the RGS instrument \citep{denherder2001} was operated in multipointing mode. Point source photons of a given energy are always recorded in the same pixel of the detector due to the pointing stability of XMM-{\em Newton}. This means that if bad pixels in the detectors coincide with spectral features of interest, these may be lost. Multipointing mode uses five different pointings with offsets in the dispersion direction. In this way, the bad pixels, which often hamper the analysis of narrow spectral features, fall at a different energies for each pointing. Combining the spectra allows us to recover the true spectrum, at the expense of a slightly lower signal to noise ratio at the position of the bad pixel.\\
\indent In the analysis, we took care of manually selecting the stable orbit portion for each of the pointings. This step is not currently handled automatically by the SAS pipeline. In practice, an additional good-time-interval table was used, in addition to the background event-filter, to select the events belonging to the stable part of the orbit. This led to a loss of only $\sim 1$ ks of effective exposure time per orbit. 
Furthermore, observation 801 was affected by episodes of background flaring. We selected only the events with background count rate less than $0.2\, \text{cts s}^{-1}$, as recommended by the SAS\footnote{https://www.cosmos.esa.int/web/xmm-newton/what-is-sas} guidelines. The background filtering led to a loss of about 21\,ks. 
The total net exposure time (including both orbits) is then 258\,ks with $123497\pm388$ net source counts.    
\section{Spectral analysis}
\label{section:3}
\begin{figure}
	\includegraphics[width=\columnwidth]{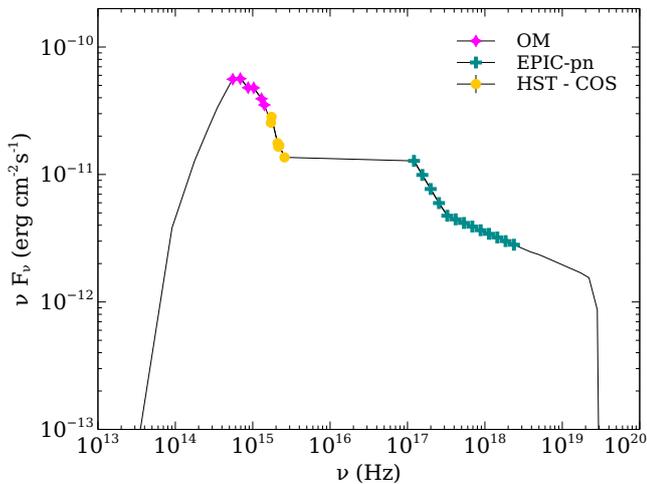}
    \caption{Adopted spectral energy distribution of I~ZW~1 based on the time-averaged simultaneous observations from XMM-{\em Newton} and HST in 2015.}
    \label{fig:sed}
\end{figure}

\begin{figure*}
	\includegraphics[width=0.85\textwidth]{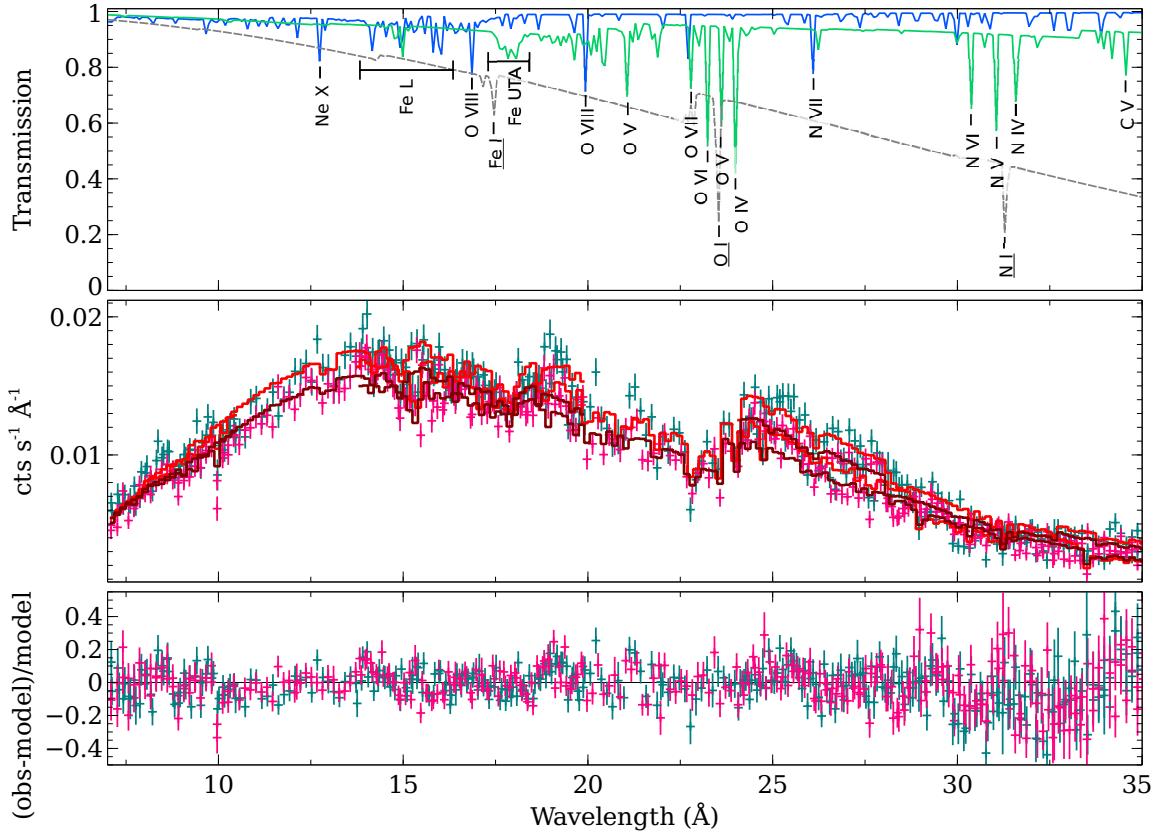}
    \caption{Best-fit parameters for the combined fit of the 2015 observations. The upper panel shows the transmission spectra of the two warm absorber models. The low-ionization component (component 1) is shown in green and the high-ionization component (component 2) is shown in blue. The strongest absorption features are labelled. The grey dashed-line refers to Galactic absorption and the features it produces are underlined. The middle panel shows the data (magenta, green) and the best-fit model (dark red, red) for observations 301 and 801 respectively. Data from both RGS instruments are presented. The lower panel shows the respective residuals.}
    \label{fig:best_fit_2xabs}
\end{figure*}

The absorption-line rich soft X-ray spectra obtained from RGS during the two observations were analysed using the fitting package SPEX v.3.02 \citep{kaastra1996}. SPEX photoionization absorption model \textsc{xabs} \citep{steenbrugge2003} is able to model the complex absorption features observed in the spectra taking into account all relevant ions in a consistent manner. The \textsc{xabs} model calculates the transmission of a slab of material in photoionization equilibrium, by interpolating over a fine grid of column density $N_{\text{H}}$ and ionization values $\text{log}\ \xi$. The ionization balance is given as an input to \textsc{xabs} and is calculated with CLOUDY v.13.01 \citep{ferland2013} using the spectral energy distribution (SED) specific for these observations, for which the ionizing luminosity is calculated between 1 - 1000 Ryd.\\
\indent The SED was constructed by using the simultaneous observations of I~ZW~1 in 2015, taken with XMM-{\em Newton} and HST. The shape of the broad X-ray ionizing continuum (0.5 - 10 keV) is obtained by fitting the time-averaged spectrum of the EPIC-pn camera on board XMM-{\em Newton}. This phenomenological fit consists of a broken-power law with $\Gamma_{1}\sim3.02$, $\Gamma_{2}\sim2.26$, and an energy spectral break at $\sim 1.4\ \text{keV}$. The unabsorbed (both of Galactic and local ionized absorption) continuum is then used to construct the SED. The UV data refer to simultaneous observations with the COS instrument on HST, while the XMM optical monitor (OM) extends the data to the optical band. Both the COS and OM data were corrected for the effects of interstellar extinction. At longer wavelengths the SED was completed by making use of the default AGN continuum in CLOUDY \citep{mathews1987}. The continuum was extended above 10 keV, with a cutoff at $\sim150$ keV. The broadband SED is presented in Fig.~\ref{fig:sed}.\\
\indent The RGS energy range is limited, and thus we opted to fit a phenomenological continuum model that describes the data well in this range. The continuum shape for each observation was best described by a broken power-law (see Table \ref{table:continuum_parameters}). We have then combined the two observations in sectors\footnote{More information on how to create different sectors to analyse several observations simultaneously in SPEX can be found at http://var.sron.nl/SPEX-doc/cookbookv3.0/cookbook.html}. In this way we are able to fit the spectra of the two observations simultaneously, increasing the signal to noise ratio to constrain the warm absorber parameters, but still allowing continuum spectral changes between observations. Throughout the analysis, we use the C-statistic as shown in \cite{kaastra2017}, and the optimal data bin size, which rebins the data taking into account the signal to noise ratio as well as the instrumental resolution \citep[see][for details]{kaastra&bleeker2016}. The optimal bin size can be achieved with the command \textsc{obin} in SPEX.\\
\indent The residuals left from fitting the continuum (C-stat/$d.o.f.\sim4737/3723$) clearly indicate absorption features, which were already identified in the past as two distinct warm absorber components \citep{costantini2007}.
\begin{table}
  \centering
  \caption{Best-fit parameters for the underlying continuum of the RGS spectra. Fluxes are in units of $10^{-12}\ \text{erg s}^{-1}\ \text{cm}^{-2}$, unabsorbed luminosities in units of $10^{43}\ \text{erg s}^{-1}$.}
  \begin{tabular}{lcccr} 
		\hline
		 Param. & Obs. 301 & Obs. 801\\
		\hline
		 $\Gamma_{1}$& $2.93\pm0.01$ & $2.98\pm0.01$\\ 
		 $\Gamma_{2}$& $2.52\pm0.1$ & $2.47\pm0.1$\\
		 $E_\text{break} \text{(keV)}$ & $1.4\pm0.1$ & $1.4\pm0.1$\\
		 $F_{0.5-2\ \text{keV}}$ & $6.64\pm0.07$ & $7.39\pm0.09$ \\
		 $L_{0.5-2\ \text{keV}}$ & $8.54\pm 0.09$ & $9.5\pm0.1$ \\
		\hline
	\end{tabular}
	\label{table:continuum_parameters}
\end{table}
\begin{table}
  \centering
  \caption{Best-fit parameters for the combined fit of the 2015 observations. Components 1 and 2 correspond to the low and high-ionization phases, respectively.}
  \begin{tabular}{lcccr} 
		\hline
		 & $N_\text{H}$ & $\text{log}\ {\xi}$ & Velocity & $\sigma_\text{rms}$\\
		& ($\times 10^{20} \text{cm}^{-2})$ & & ($\text{km s}^{-1}$) & ($\text{km s}^{-1}$) \\
		\hline
		Comp 1 & $3.4\pm0.4$ & $-0.23\pm0.06$ & $-1870\pm70$ & $70\pm10$\\
		Comp 2 & $9\pm2$ & $1.96\pm0.05$  & $-2500\pm100$ & $30\pm10$\\
		\hline
	\end{tabular}
	\label{table:2xabs}
\end{table}
Adding a photoionization absorption model to the fit resulted in an improvement for which we reached C-stat/$d.o.f\sim4612/3719$. This model accounts for strong absorption features observable around 21 - 25 $\AA$, due to multiple oxygen transitions (\ion{O}{v} - \ion{O}{vii}), as well as for the iron unresolved transition array (UTA). Finally, an additional photoionization absorption model is required to fit residuals at shorter wavelengths (10 - 17 $\AA$) such as the \ion{O}{viii} edge and the Fe L complex, which results in C-stat/$d.o.f.\sim 4565/3715$. Our best-fit model to the combined RGS spectra thus confirms the presence of two distinct photoionized components, a low and a high-ionization phase, see Fig. \ref{fig:best_fit_2xabs}. Component 1, the low-ionization phase ($\text{log}\ \xi\sim-0.2$) has a column density $N_\text{H}\sim 3.4\times10^{20}\text{cm}^{-2}$ and an outflow velocity $v_{\text{out}}\sim 1900 \  \text{km} \ \text{s}^{-1}$. The high-ionization phase ($\text{log}\ \xi\sim2$, hereafter component 2) has a column density $N_\text{H}\sim 9\times10^{20}\text{cm}^{-2}$ and an outflow velocity $v_{\text{out}}\sim 2500 \  \text{km} \ \text{s}^{-1}$. The root mean square width of the absorption lines was constrained by our fit to be $\sigma_\text{rms}\sim70\ \text{km s}^{-1}$ for component 1 and $\sigma_\text{rms}\sim30\ \text{km s}^{-1}$ for component 2. The best-fit parameters for the continuum are presented in Table~\ref{table:continuum_parameters} and the best-fit parameters for the warm absorber components are presented in Table \ref{table:2xabs}. While fitting the spectrum only these four parameters for each \textsc{xabs} model were free to vary. We fixed the line-of-sight covering factor of the gas to unity. Looking at the residuals, there could be some off-set from the best-fit model, particularly at low energies. This would likely be due to the complexity of the of the continuum which in our model is represented in a simplified manner. Our analysis did not reveal the presence of additional absorber components.
\subsection{Time-resolved spectroscopy}
To investigate if there are any changes to the warm absorber parameters when considering the source's short-term time-variability, we performed a time-resolved analysis of the spectra of I~ZW~1.\\

\begin{figure}
	\includegraphics[width=\columnwidth]{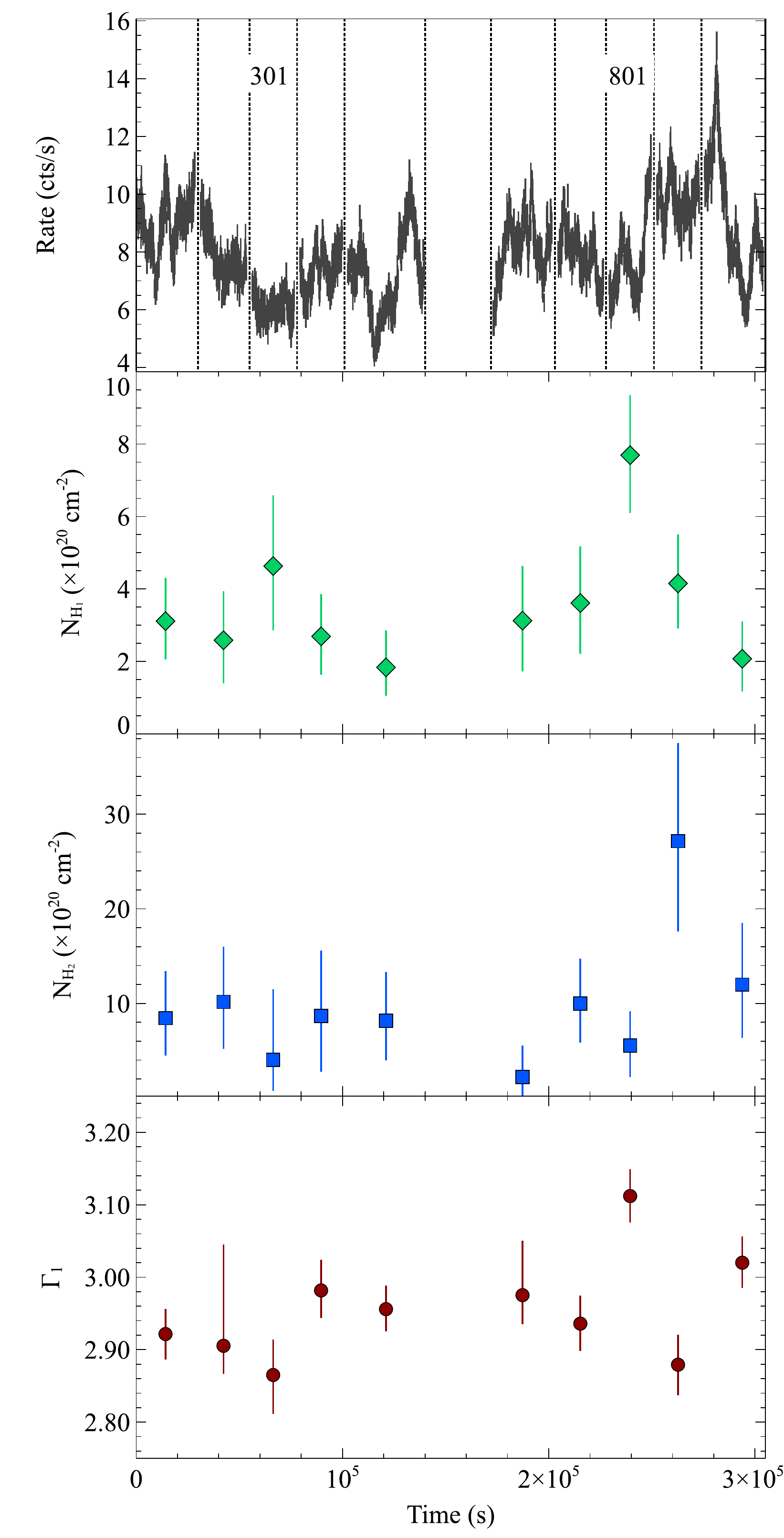}
    \caption{Best-fit parameters for the individual pointings. The behaviour of $N_\text{H}$ with time is shown for each one of the individual pointings. The ionization fixed of each of the gas components was fixed in these fits to the best-fit value we found while fitting the long segments. For reference, the photon spectral index, $\Gamma_1$, of each individual pointing is also shown (see bottom panel).}
    \label{fig:individual_pointings}
\end{figure}

We have considered at first two time segments per each orbit. Combining individual pointings ensures both the elimination of bad pixels and a fair signal-to-noise ratio, which allowed us to simultaneously fit for the ionization and column density of the gas over these timescales. We probe here timescales of $\sim$ 50-80 ks. The first segment of observation 301 is the combination of the first two pointings, while the second segment is the combination of the remaining three pointings. As for observation 801, the first three pointings were combined to create the first segment, and the remaining two pointings constitute the second segment (see top panel of Fig.~\ref{fig:individual_pointings}). The rationale behind this combination in particular relies on an attempt to combine, and analyse, similar flux levels, while ensuring a high signal-to-noise ratio. To fit the spectra we have taken advantage of the constraints we obtained when fitting the two observations simultaneously, as seen in section \ref{section:3}. We started by using the best-fit model to the combined fit of the 2015 observations (see Table~\ref{table:2xabs}), and fixing the warm absorber parameters. We then first fitted for the continuum by allowing the normalization of the power-law, the power-law indexes and the spectral break energy to vary. Subsequently we fitted the warm absorber parameters of components 1 and 2, leaving the column density and ionization of the gas as free parameters, while the continuum parameters remained thawed. The outflow velocity was kept fixed during the fits as well as the width of the absorption lines, $\sigma_\text{rms}$ to the values from Table \ref{table:2xabs}. Both these parameters can at best be estimated in the time-averaged spectrum. The best-fit parameters are listed in Table \ref{table:long_segments}.\smallskip \\
  \begin{table*}
  \centering
  \caption{ Best-fit parameters of the long time segments. The count rate refers to the RGS spectra.}
  \label{table:long_segments}
  \begin{tabular}{lccccccccr} 
		\hline
		 & $N_{\text{H, Comp 1}}$ & $N_{\text{H, Comp 2}}$ & $\text{log}\ \xi_{\text{Comp 1}}$ & $\text{log}\ \xi_{\text{Comp 2}}$ & $\Gamma_1$ & $\Gamma_2$ & $E_\text{break}$ & Rate & C-stat/\\
		& $(\times 10^{20} \text{cm}^{-2})$ & $(\times 10^{20} \text{cm}^{-2})$ & & & & & (keV) & (cts/s) & $d.o.f.$\\
		\hline
		Seg 1 & $3\pm1$ & $10\pm4$ & $-0.3\pm0.1$ & $1.9\pm0.1$ & $2.90\pm{0.03}$ & $2.6\pm{0.2}$ & $1.3\pm{0.1}$ &  $0.518\pm0.002$ & $2612/2272$  \\
		Seg 2 & $3\pm1$ & $7\pm3$ & $-0.4\pm0.2$ & $2.0\pm0.1$ & $2.97\pm{0.03}$ & $2.4\pm{0.3}$ & $1.4\pm{0.1}$ & $0.444\pm0.003$ & $4104/3458$ \\
		Seg 3 & $5\pm1$ & $4\pm4$ & $-0.3\pm0.1$ & $1.8\pm0.2$ & $3.01\pm{0.03}$ & $2.7\pm{0.1}$ & $1.2\pm{0.1}$ &  $0.476\pm0.002$ &$3046/2275$ \\
		Seg 4 & $3\pm1$ & $19\pm6$ & $-0.2\pm0.1$ & $2.00\pm0.06$ & $2.97\pm{0.03}$ & $2.4\pm{0.3}$ & $1.3\pm{0.1}$ & $0.584\pm0.004$ & $2484/2325$\\
		\hline		
	\end{tabular}
\end{table*}
Regarding the ionization state of the gas, both components do not seem to vary within the errors (see Table~\ref{table:long_segments}). Thus, it is not possible to assess whether the gas is in equilibrium with the ionizing continuum for the considered timescales or if it is able to respond to the flux variations during these observations, since such changes would not be possible to be detected with the present statistics. Variability in the column density of the warm absorber components is formally not detected at these timescales, although component 2 appears to increase in column density during the second observation. Motivated by these results, we attempted to fit the individual pointings separately and investigate further possible column density variations on shorter timescales.\\

We thus fit, in the same manner as described above, the spectrum of each individual pointing separately to investigate possible variations on shorter timescales, and if such variability would directly relate to the intrinsic changes in the flux of the source. The best-fit parameters can be found in Table \ref{table:individual_pointings}.\\
  \begin{table*}
  \centering
  \caption{ Best-fit parameters of the individual pointings. The count rate refers to the individual RGS spectra.}
  \label{table:individual_pointings}
  \begin{tabular}{lcccccccr} 
		\hline
		& $N_{\text{H, Comp 1}}$ & $N_{\text{H, Comp 2}}$ & $\Gamma_1$ & $\Gamma_2$ & $E_\text{break}$ & Rate & C-stat/\\
		& ($\times 10^{20} \text{cm}^{-2})$ & ($\times 10^{20} \text{cm}^{-2})$ & & & (keV) & (cts/s) & $d.o.f.$\\
		\hline
		301-1 & $3\pm1$ & $8\pm5$ & $2.92\pm{0.03}$ & $2.4\pm{0.2}$ & $1.3\pm{0.1}$ & $0.527\pm0.004$ & 1703/1540 \\
		301-2 & $3\pm1$ & $10\pm5$ & $2.91^{+0.1}_{-0.03}$ & $2.6\pm{0.4}$ & $1.4\pm{0.6}$ & $0.468\pm0.003$ & 3217/2610 \\
		301-3 & $5\pm2$ & $4_{-3}^{+7}$ & $2.87\pm{0.05}$ & $2.6^{+0.2}_{-0.5}$ &  $1.3\pm{0.2}$ & $0.358\pm0.003$ & 1773/1513 \\
		301-4 & $3\pm1$ & $9\pm7$ & $2.98\pm{0.04}$ & $2.4^{+0.4}_{-0.7}$ & $1.5\pm{0.2}$ & $0.444\pm0.003$ & 1832/1612  \\
		301-5 & $2\pm1$ & $8\pm5$ & $2.96\pm{0.03}$ & $2.2\pm{0.4}$ & $1.5\pm{0.1}$ &  $0.445\pm0.003$ & 1988/1720\\
		801-1 & $3\pm1$ & $2\pm2$ & $2.98^{+0.07}_{-0.04}$ & $2.3\pm{0.2}$ & $1.2\pm{0.5}$  & $0.455\pm0.003$ & 3101/2646 \\
		801-2 & $3\pm1$ & $10\pm5$ & $2.94\pm{0.04}$ & $-$ & $-$ & $0.455\pm0.003$ & 3165/2745 \\
		801-3 & $8\pm2$ & $ 6\pm3$ & $3.11\pm{0.04}$ & $-$ & $-$  & $0.456\pm0.003$ & 1967/1708 \\
		801-4 & $4\pm1$ & $27\pm10$ & $2.89\pm{0.04}$ & $2.6\pm{0.9}$ & $1.4\pm{0.1}$  & $0.578\pm0.004$ & 1889/1731 \\
		801-5 & $2\pm1$ & $12\pm6$ & $3.02\pm{0.04}$ & $1.9\pm{0.2}$ & $1.3\pm{0.1}$  & $0.550\pm0.004$ & 1781/1648 \\
		\hline
	\end{tabular}
\end{table*}
We firstly note that it is indeed difficult to constrain the warm absorber parameters for each of the individual pointings. The lower signal-to-noise ratio does not allow for reliable constraints on the ionization parameters of the two gas components. To enable us to constrain the changes in the column density more accurately, we have fixed the ionization parameter of each of the gas components to the best-fit value we found in the previous step (see Table~\ref{table:long_segments}). The results are presented in Table~\ref{table:individual_pointings} and Fig.~\ref{fig:individual_pointings}. In Fig.~\ref{fig:individual_pointings}, we have also added the best-fit values for the photon spectral index, $\Gamma_1$, of the underlying continuum of each individual spectra, for reference. As can be seen in Table \ref{table:individual_pointings}, the underlying continuum model, which consists of a broken power-law, is mostly influenced by $\Gamma_1$ over the energy range covered by RGS, as the energy of the spectral break $E_\text{break}$ can confirm. $\Gamma_2$ only influences a limited part of the spectra ($<10\ \AA$), where no relevant warm absorber features are found. According to the results of the fits to the spectra of each individual pointing, the column density of the two gas components does not appear to vary at a significant level. The apparent increase in the column density of component 2 towards the end of the second observation lies at most at $2\ \sigma$ from the weighted mean. We however note that the large uncertainties on the estimation of $N_\text{H}$, due to the low signal-to-noise ratio of each individual spectra, do not allow us in any case to determine if any intrinsic variations are present.

\section{Discussion}
\label{section:4}
These new observations confirm the presence of two ionized absorbing gas components in the soft X-ray spectrum of I~ZW~1. In this section we discuss our results in the light of previous observations and their implications in the context of warm absorber models and geometries. 
\subsection{A two component warm absorber}
\begin{figure}
	\includegraphics[width=\columnwidth]{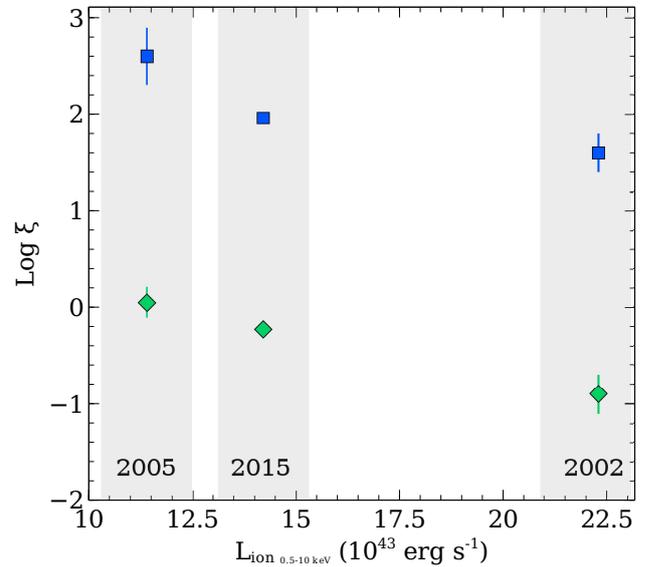}
    \caption{X-ray luminosity versus ionization parameter. The ionization state of both gas components do not correlate with the ionizing luminosity. Note also the similar behaviour of both gas components, suggesting the two gas phases appear to be linked. Notice that the ionization states of the gas is represented here over a broad range to accommodate both components. The changes in the ionization state of the gas throughout the years can be better appreciated in Table \ref{table:history_wa}. Additionally, please note that at this scale the error bars in the estimated values for the ionization parameter, $\xi$, of components 1 and 2 in 2015 are within the size of the marker (see table \ref{table:history_wa}).}
    \label{fig:l_vs_xi}
\end{figure}

\begin{figure}
	\includegraphics[width=\columnwidth]{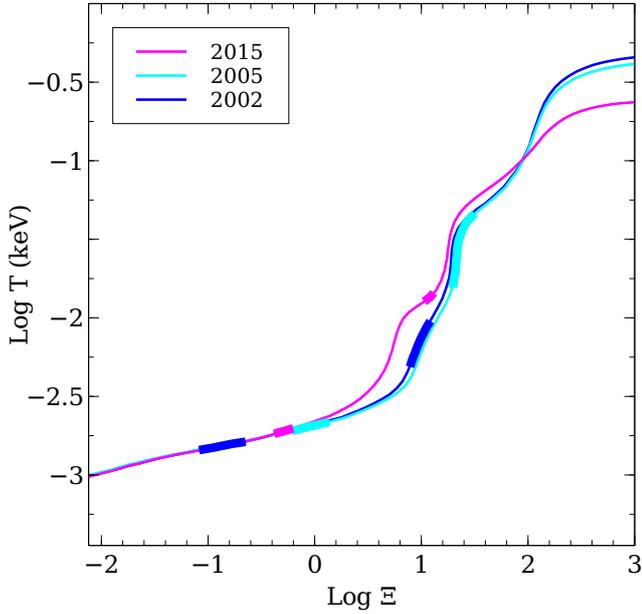}
    \caption{Stability curves for the different observations of I~ZW~1 by XMM-{\em Newton}. The low and high-ionization warm absorber components are marked in the curves by the thicker lines. }
    \label{fig:s_curve}
\end{figure}

Our results are particularly interesting when placed in context with previous observations of I~ZW~1 (See Table~\ref{table:history_wa}). I~ZW~1 was firstly observed by XMM {\em Newton} in 2002 \citep[see][]{gallo2004}. When presenting the new soft X-ray data from 2005, \cite{costantini2007} characterized the warm absorber in both epochs by performing an analysis similar to this work, also using the \textsc{xabs} model available in SPEX. \cite{costantini2007} found that the soft X-ray spectra during both epochs required two warm absorber models, including low and high-ionization components, with similar column densities. While the column densities did not change dramatically over those years, there was a variation of the ionization parameters of both components that oddly suggested an anti-correlation between the X-ray luminosity and the ionization state of both components of the gas. In these lower exposure observations, the signal-to-noise ratio was not sufficient to determine the outflow velocity of the high-ionization warm absorber component. The low-ionization component shows outflow velocities in 2002 and 2005 compatible to the outflow velocity observed in 2015. More interestingly, the low-ionization component was found to likely have a counterpart in the UV \cite[see][]{laor1997,costantini2007}. This component has been found to have the same outflow velocity since it was first observed in the UV \citep{laor1997}. In \cite{costantini2007} an intrinsic neutral absorber was also identified. We find, in our analysis, that the improvement on the Galactic absorption measurements, which now include both the atomic and the molecular hydrogen components \citep{willingale2013}, is likely be the reason why our fits do not require an extra neutral absorber. Furthermore, in these longer exposure data, we do not identify an \ion{O}{i} edge at the redshift of the source, which would be present in the case of neutral absorption.\\
  \begin{table}
  \centering
  \caption{Recent history of the X-ray WA in I~ZW~1}
  \begin{tabular}{lcccr} 
		\hline
		& & $N_\text{H}$ & $\text{log}\ {\xi}$ & Velocity\\
		& & ($\times 10^{20} \text{cm}^{-2})$ & & ($\text{km} \text{s}^{-1}$)\\
		\hline
		Comp 1 & 2002 & $24\pm5$ & $-0.9\pm0.2$ & $-1800\pm400$\\
		& 2005 & $13\pm3$ & $0.05\pm0.16$  & $-1700\pm400$\\
		& 2015 & $3.4\pm0.4$ & $-0.23\pm0.06$ & $-1870\pm70$\\
		\hline
		Comp 2 & 2002 & $13\pm4$ & $1.6\pm0.2$ & -\\
		& 2005 & $13\pm5$ & $2.6\pm0.3$  & -\\
		& 2015 & $9\pm2$ & $1.96\pm0.05$  & $-2500\pm100$\\		
		\hline
	\end{tabular}
	\label{table:history_wa}
\end{table}
\begin{figure*}
	\includegraphics[width=0.65\textwidth]{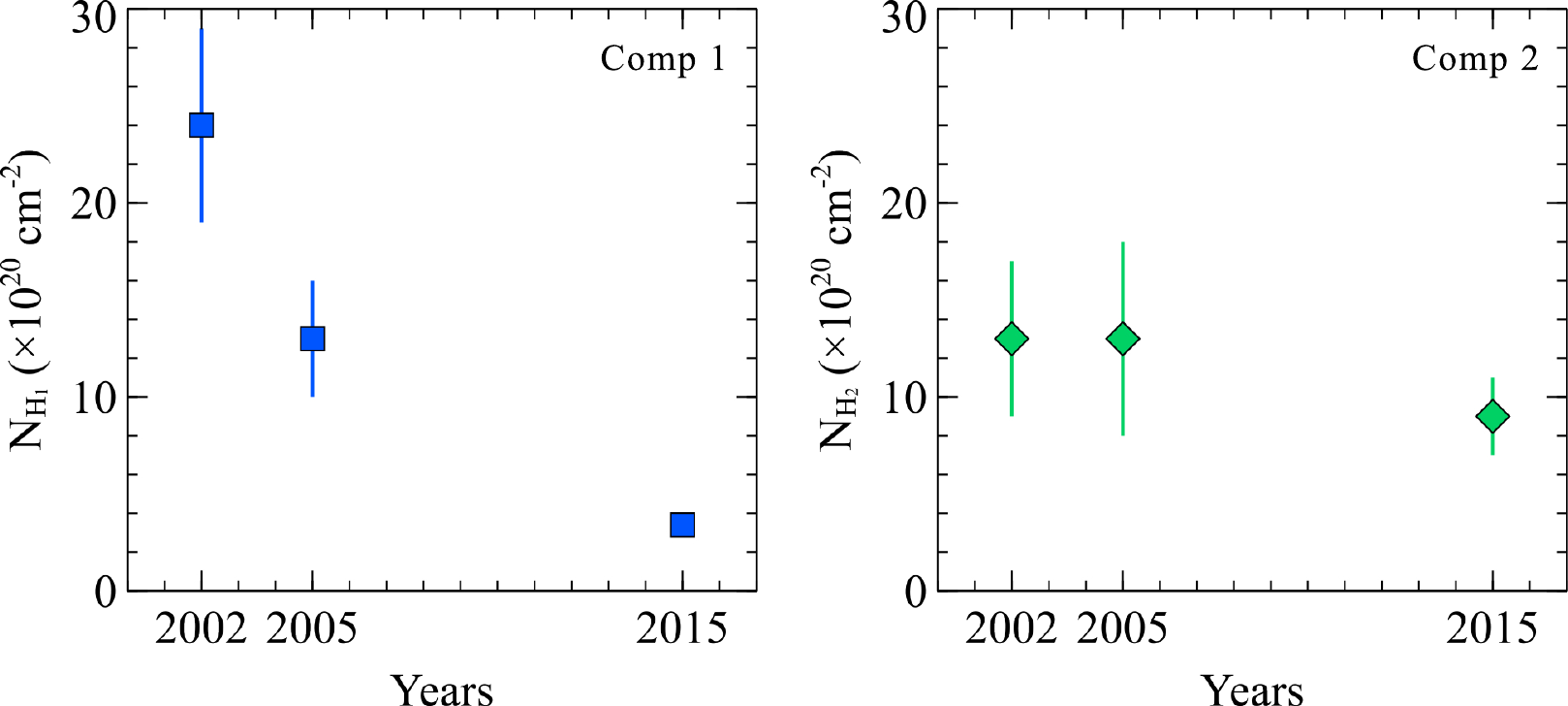}
    \caption{$N_{\text{H}_1}$ and $N_{\text{H}_2}$ behaviour throughout the years, for components 1 and 2 respectively. Please note that at this scale the error bars in the estimated value for the column density of component 1 in 2015 are within the size of the marker (see table \ref{table:history_wa}).}
    \label{fig:nh_vs_time}
\end{figure*}
Our observations show that the ionization state of the gas is lower for both components when compared to the last observations, even though the X-ray luminosity is higher in 2015 than it was in 2005 (see Table \ref{table:history_wa}). Fig. \ref{fig:l_vs_xi} shows the X-ray luminosity in each of the epochs versus the estimated ionization parameter for each of the absorber components. The X-ray luminosity alone is a good diagnostic for ionizing luminosity in this case, since the X-ray ions are more sensitive to the X-ray photons rather than to the UV continuum \citep[e.g.][]{netzer1996}. When photoionization occurs, a variable source will have an impact on the ionization balance of the gas. For an equilibrium situation, the gas responds instantaneously to such changes, becoming more ionized as the flux increases and recombining when the flux drops. In the presence of a low density gas (and depending on its distance to the continuum source), the response to changes in luminosity may be delayed which results in a complex time-dependent behaviour \citep{nicastro1999,silva2016}. Remarkably, I~ZW~1 shows instead an apparent anti-correlation with X-ray luminosity on timescales of years. This could derive from a scenario in which the outflowing absorbing gas is in constant non-equilibrium with its ionizing source or, could result instead from the presence of transiting gas, crossing our line of sight. \\
\indent In either case, the two warm absorber components must be linked. The two show the same ionization behaviour regardless of the luminosity and similar outflowing velocities. To co-exist, the two components are expected to be in pressure equilibrium by classical warm absorber scenarios (see more in section \ref{sec:scenarios}). To investigate if the low and high-ionization phases of the absorbing gas are in pressure equilibrium, we have generated the thermal stability curves for I~ZW~1 by plotting the pressure ionization parameter, $\Xi$ as a function of the electron temperature, $T$, see Fig. \ref{fig:s_curve}. The pressure ionization parameter is defined as $\Xi=L/4\pi r^{2}cp=\xi/4\pi ckT$, where $c$ is the speed of light, $p$ is the pressure, $k$ is the constant of Boltzmann, and $T$ is the electron temperature. We computed with CLOUDY the corresponding electron temperature for a grid of ionization parameters $\xi$, thus allowing us to estimate $\Xi$. The two ionized gas components are overplotted on the stability curves for each epoch. To be in pressure equilibrium both components would need to share the same $\Xi$. As it is clear from Fig. \ref{fig:s_curve}, this is not the case for the two gas phases, as they lie far apart in pressure ionization parameter.  An alternative ways to sustain the co-existence of both components which does not require pressure equilibrium is a scenario of  radiation pressure confinement \citep{stern2014} or magnetic confinement \citep{rees1987}.\\
\indent Furthermore, the column density of the low ionization component has dropped by at least a factor of three, since I~ZW~1 was last observed in 2005 (see Fig. \ref{fig:nh_vs_time}). Meanwhile, the outflow velocity of this component has remained constant for the past 20 years, when its UV counterpart was first observed.

Finally, we have also investigated the short-term behaviour of the warm absorber. Our analysis does not show any significant variations in the ionization state and column density of the outflow components in timescales of $\sim$ hours. The lower signal-to-noise ratio of the individual spectra, however, may confuse the detection of true intrinsic variations.

\subsection{The origin of the warm absorber in I~ZW~1}
\label{sec:scenarios}
\begin{figure}
	\includegraphics[width=\columnwidth]{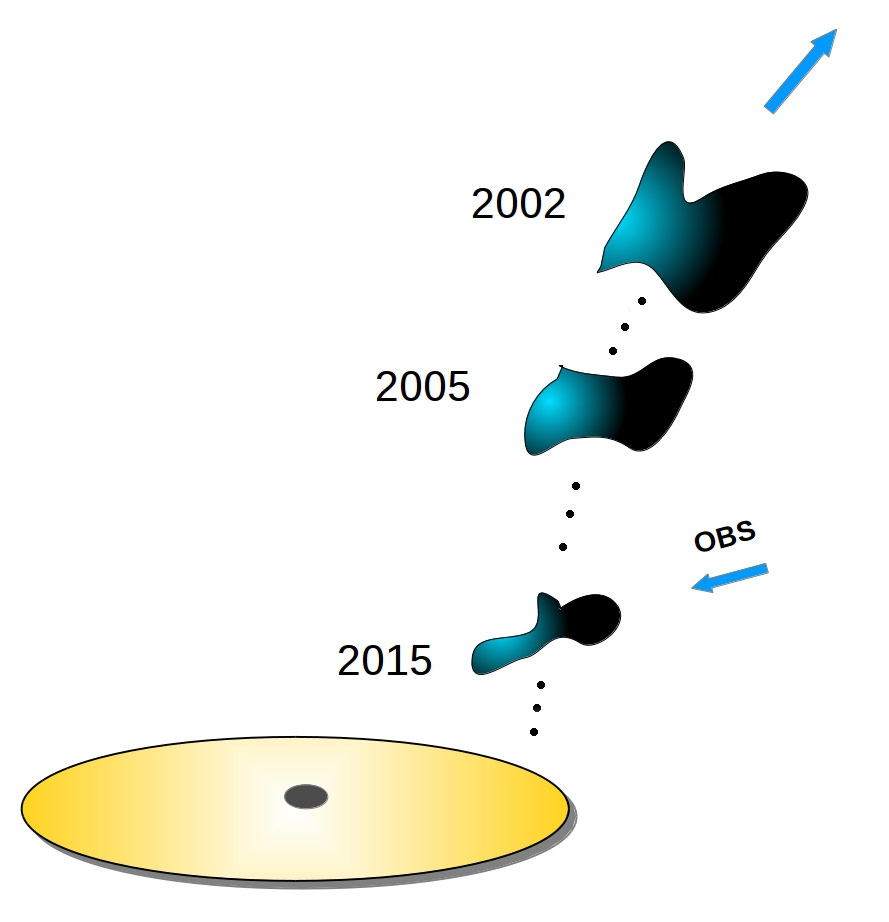}
    \caption{Schematic view of the phenomenological model for the warm absorber in I~ZW~1. In this scheme, we highlight only the events we witnessed on the occasion of the observations with XMM-{\em Newton} throughout the years.}
    \label{fig:agn_outflow_clumpy}
\end{figure}
Constraints on the location of the X-ray warm absorber are generally derived once the density of the gas is estimated. Estimating the density of the gas is only possible when the photoionized gas responds to changes in the ionizing continuum, allowing for a recombination timescale to  be measured \citep[see e.g.][]{krongold2007}. In I~ZW~1, instead of the expected linear response, we observe the ionization state of both gas components, while varying together, to be uncorrelated with the ionizing luminosity along the years and as such, photoionization equilibrium does not easily apply.\smallskip \\
\indent We can derive a crude estimate for the location of the warm absorber by assuming a spherical outflow with a uniform filling factor $f$, and a density that decreases with radius, $n(R)$. Following \cite{blustin2005}, we assume that all the warm absorber mass is contained within a thin layer of thickness $\Delta r$. The thickness of such a layer must be smaller than or at most equal to the distance of the warm absorber to the central source such that
\begin{equation}
\label{eqn:thickness_r}
\frac{\Delta r}{R} \le 1.
\end{equation}
The column density along our line of sight, for a specific ionization parameter $\xi$, can be expressed as a function of the gas density $n(R)$, the thickness of the layer $\Delta r$ and the volume filling factor $f$ as
\begin{equation}
\label{eqn:column_density}
N_\text{H}\sim n(R) \Delta r f.
\end{equation}
Since the ionization parameter in the shell is given by
\begin{equation}
\label{eqn:xi_shell}
\xi=\frac{L_\text{ion}}{n(R) R^2},
\end{equation}
we get the maximum distance of the warm absorber to the central source

\begin{equation}
\label{eqn:r_max}
R\le \frac{L_\text{ion} f}{\xi N_\text{H}}.\smallskip
\end{equation}

\indent The volume filling factor $f$ can be estimated by equating the momentum of the outflow to the momentum of the radiation it absorbs plus the momentum of the radiation it scatters \citep{blustin2005}. Using the parameters from the best-fit of the averaged 2015 spectrum, we have calculate the volume filling factor in I~ZW~1 to be $f_1\sim4\times10^{-5}$ and $f_2\sim3\times10^{-3}$, which respectively locates the warm absorber within 45 pc for component~1 and 9 pc for component~2.\smallskip \\
\indent A minimum radius for the location of the warm absorber can also be estimated by assuming the outflow velocity $v_\text{out}$ is greater or equal to the escape velocity such that

\begin{equation}
\label{eqn:r_min}
R\ge \frac{2GM_\text{BH}}{v^{2}_\text{out}}.
\end{equation}
This assumption would put the minimum location of the warm absorber at a distance of $R_1\ge0.07\ \text{pc}$ from the central source for component 1 and $R_2\ge0.04\ \text{pc}$ for component 2.\\

In classical warm absorber models, the gas would be expected to respond to variations in the ionizing continuum. For thin spherical shells of material, the outflow is radially stratified, with the ionization of the gas being mostly dependent on its distance to the central source \citep[see e.g.][]{steenbrugge2005}. The long term variability observed on I~ZW~1 suggests instead the density of the gas to be the main driver of the ionization. Alternatively, some studies have suggested a clumpy outflow in which tenuous hot, high-ionization, gas absorbs the X-rays, while surrounding discrete filaments of cold, low-ionization, UV absorbing gas. This scenario allows for the two components to be co-located and could explain the observed variability, but requires both phases to be in pressure equilibrium \citep{krolik&kriss2001}. However, the high and low-ionization components observed in I~ZW~1 are not in pressure equilibrium (see Fig. \ref{fig:s_curve}), and as such, this scenario also fails to be a candidate for the possible geometry of this outflow. \\
\indent The peculiar characteristics observed for the warm absorber in I~ZW~1 cannot be explained through classical warm absorber scenarios. To start with, changes in the ionization state of the gas throughout the years do not appear to be linked to variations in the ionizing luminosity as is expected for photoionization. Variations in the column density of the gas are also observed on long timescales. Furthermore, these variations in ionization and opacity seem to be correlated for both components, which also happen to be outflowing at similar velocities. The outflow velocity of the low-ionization component, in particular, seems to be unchanged for the last 20 years, since its UV counterpart was first observed. We propose here an alternative geometrical model that could accommodate the apparent oddities of this source.\\
 \indent Due to the persistent outflow velocity observed in the colder component for the past two decades, we suggest that we are observing of the same ejection phenomenon throughout all the observations of I~ZW~1.
In this picture, the gas identified in the UV/X-rays, the low-ionization component, primarily constitutes the flow. The skin layer facing the source would naturally become highly ionized, resulting in the observed high-ionization component which is detected only in the X-rays. Radiatively driven wind models have shown in the past a two-dimensional flow structure similar to what is here observed, where a dense, slower outflow is confined at the polar side by a less dense higher velocity stream \citep{proga1998}. This could explain the observed ionization and kinematic properties of the warm absorber in I~ZW~1, which suggest the high-ionization, hence lower density, phase of the flow to be faster than the low-ionization component ($v_\text{Comp 1}\sim-1900\ \text{km}\ \text{s}^{-1}$, $v_\text{Comp 2}\sim-2500\ \text{km}\ \text{s}^{-1}$). This connection between the two components would also offer a natural explanation for the correlated variability observed on long timescales regarding the ionization state and opacity of both gas components. In fact, preliminary analysis of the UV data belonging to this multiwavelength campaign (Giustini et al. in prep.) has already shown line-locking to be at play, a signature of a radiatively driven wind, pointing to radiation pressure as the likely driving mechanism of the outflow \citep{proga2000}. Furthermore, the geometrical model we propose (see Fig. \ref{fig:agn_outflow_clumpy}), considers the outflow to be clumpy. Different clumps, having different densities, and thus different ionization, cross the observer's line of sight at different epochs, resulting in the observed changes over the years, both in ionization and column density. In particular, the ionization state of the gas at different epochs is driven by different densities of the clumps and not by changes in the ionizing luminosity affecting gas components at different radii, as is usually assumed. This is in agreement with recent findings in 3D simulations of line-driven winds whose results show the presence of clumps. The clumpiness of the flow may be responsible for alterations in the ionization state of the gas since, in this case, it dependents on the density of the gas along a single line of sight and can be affected by single over/under densities \citep{dyda2018}. This would explain the non-trivial behaviour of the ionization of the gas as a function of luminosity over long timescales. Also the changes in opacity for the different observations (2015, 2005 and 2002) can be easily understood as a natural consequence of a clumpy outflow.\\
The driving radiation force can also compress the gas resulting in a scenario of radiation pressure confinement \citep{stern2014}, which naturally explains a multiphase outflow without the need for pressure equilibrium between phases since the pressure increases significantly throughout the slab, with decreasing ionization. In a radiation pressure confinement scenario, the radiation pressure compresses the gas instead of accelerating it, which results in an outflow with constant velocity, as observed in I~ZW~1. Alternatively, magnetic confinement could also allow for a co-existence of the two warm absorber phases even if these are not in pressure equilibrium \citep{rees1987}.
\section{Conclusions}
\label{section:5}
We have performed a detailed analysis of the recent observations by RGS on board XMM-{\em Newton} of the narrow-line Seyfert~1 galaxy I~ZW~1. The extensive modelling of the observed spectral features through time-averaged spectral fitting and time-resolved spectroscopy, we conclude the following:
\begin{enumerate}
\item We have confirmed the absorbing gas to be composed of two ionization phases, as previously reported by \cite{costantini2007} based on the historical data of I~ZW~1. The long-lived low-ionization component (log $\xi\sim0$) has been associated with the observed absorbing gas in the UV and its outflow velocity has remained unchanged for nearly 20 years. The high-ionization component (log $\xi\sim2$) shows a slightly higher outflow velocity than the low-ionization component ($\sim-2500 \ \text{km s}^{-1}$ and $\sim-2000\  \text{km s}^{-1}$ respectively) and the long term variability of the gas suggests the two components to be linked.
\item The long-term variability of I~ZW~1 disagrees with the commonly assumed scenario of photoionization equilibrium. The low and the high-ionization phases vary together in ionization state, yet these variations do not correlate with changes in the ionizing luminosity. This suggests the density of the gas to be the main driver of the observed changes in ionization, instead of a response to the variability of the ionizing source. The column density of the low-ionization component is also variable on timescales of years. The high-ionization component is consistent with being constant at such timescales. We have also studied the variability of the warm absorber on short-timescales. Our observations do not show significant variations on timescales of hours.
\item Classical warm absorber models fail to explain the variability of the ionized gas in I~ZW~1. As such, we propose a phenomenological model in which the gas originates from a inhomogeneous clumpy outflow, possibly radiatively driven. The low-ionization component primarily constitutes the flow, while the skin layer of the clumps facing the ionizing source becomes more ionized and accounts for the high-ionization phase observed. In this scenario, the ionization of the gas is primarily dependent on the density of the clumps. Different clumps have different densities, hence the observed variability in ionization throughout the years. Our suggested explanation through this alternative geometry shows similarities with recent results derived from 3D simulations of line-driven winds \citep{dyda2018}. Furthermore, our estimates suggest the gas to be located between 0.07 pc and 9 pc, implying that the outflow may even be originating from the accretion disc.
\end{enumerate}

The non-typical behaviour of the warm absorber in I~ZW~1 demonstrates that, despite extensive studies, these systems are still not fully understood and substantiates the need for new theoretical models capable of reproducing such complex behaviour, particularly regarding the variability of the gas. Furthermore, long and, in particular, multi-wavelength observations are crucial to fully characterize the warm absorber zoo in AGN and assess the potential impact of the ionized gas to AGN feedback.
\section*{Acknowledgements}
The authors would like to thank I. Psaradaki for providing the illustration presented in Fig. \ref{fig:agn_outflow_clumpy}. C. V. Silva acknowledges support from NOVA (Nederlandse Onderzoekschool voor Astronomie). The Space Research Organization of the Netherlands is supported financially by NWO, the Netherlands Organization for Scientific Research. This work was supported by NASA through a grant for HST program number 13811 from the Space Telescope Science Institute, which is operate d by the Association of Universities for Research in Astronomy, Incorporated, under NASA contract NAS5-26555. WNB acknowledges support from Space Telescope Science Institute grant HST-GO-13811.004-A. In this work we made use of observations obtained with XMM-{\em Newton}, an ESA science mission with instruments and contributions directly funded by ESA Members States and the USA (NASA).  




\bibliographystyle{silva_costantini16}
\bibliography{cvsilva} 







\bsp	
\label{lastpage}
\end{document}